# Broadcast Approach and Oblivious Cooperative Strategies for the Wireless Relay Channel – Part I : Sequential Decode-and-Forward (SDF)

Evgeniy Braginskiy , Avi Steiner and Shlomo Shamai (Shitz)

**Abstract**

In this two part paper we consider a wireless network in which a source terminal communicates with a destination and a relay terminal is occasionally present in close proximity to the source without source's knowledge, suggesting oblivious protocols. The source-relay channel is assumed to be a fixed gain AWGN due to the proximity while the source-destination and the relay-destination channels are subject to a block flat Rayleigh fading. A perfect CSI at the respective receivers only is assumed. With the average throughput as a performance measure, we incorporate a two-layer broadcast approach into two cooperative strategies based on the decode-and-forward scheme – Sequential Decoded-and Forward (SDF) in part I and the Block-Markov (BM) in part II. The broadcast approach splits the transmitted rate into superimposed layers corresponding to a "bad" and a "good" channel states, allowing better adaptation to the actual channel conditions In part I, the achievable rate expressions for the SDF strategy are derived under the broadcast approach for multiple settings including single user, MISO and the general relay setting using successive decoding technique, both numerically and analytically. Continuous broadcasting lower bounds are derived for the MISO and an oblivious cooperation scenarios.

***Index Terms*****—**broadcast approach, collocated users, decode-and-forward (DF), layered transmission, oblivious cooperation, relay channel

## I. Introduction

The benefits of cooperation among users in wireless networks have been studied extensively from a variety of aspects including channel capacity [5],[10] cooperative diversity [11] and diversity gain [16]. While channel uncertainties (e.g fading) play an important role in these studies, topology uncertainties have gained less attention. In practical wireless networks, however, it is often difficult for each user to keep track of neighboring terminals which can potentially assist in the transmission of its information. Under such circumstances, users are forced to employ an oblivious communication protocols, which do not require the source's knowledge regarding the availability of nearby assisting terminals.

The network used to demonstrate the oblivious nature of the cooperation schemes is a relay channel [6] where the source and the relay are collocated. Works concerning the capacity of the relay channel and its

E.Braginskiy , A.Steiner and S.Shamai (Shitz) are with the Department of Electrical Engineering, Technion- IIT, Haifa , 32000, Israel. Email {bevgeniy@tx, savi@tx, sshlomo@ee}.technion.ac.il. This work was supported by the Israel Science Foundation and by the European Commission in the framework of the FP7 Network of Excellence in Wireless Communications NEWCOM++.

implications on cooperation in wireless channels include [6],[10], and [19, references therein] where large scale networks are treated. While the optimal strategy for employing relays in wireless networks is not yet understood, several approaches have been suggested in the literature, out of which we focus on the *decode-and-forward* (DF) . In the DF [1], [6, references therein] the relay first decodes the message sent by the source, and then re-encodes it and retransmits the same message to the destination. As a performance measure, we use a variant of the outage capacity [1] by considering *expected throughput* obtained by multiplying the attempted rate by the successful decoding probability.

The problem of cooperative oblivious protocols for the relay channel has been treated for various scenarios. In [1], a form of the decode-and-forward strategy was involving variable-length coding scheme was analyzed for a wireless network consisting of a remotely located source sending information to one of $K$ physically collocated users.. In [2], the setting is reduced to $K = 2$ and strategies based on *amplify-and-forward* and *compress-and-forward* with variable amounts of side information based on the Wyner-Ziv quantization are treated. In [3], several oblivious protocols for collocated source and relay are addressed, based on *decode-and-forward* and *compress-and-forward* , with the latter incorporating side information in the relay's quantization of the received signal [20].

In this paper, we incorporate the broadcast strategy [4] into the relay channel . This strategy stems from the broadcast channel [7], where a single transmission is directed to a number of receivers, each enjoying possibly different channel conditions. Since the slowly fading channel without transmitter CSI may be viewed as a compound channel with the channel realization as a parameter, it is essentially what the broadcast strategy is. Explicitly solved for the SISO and the SIMO channels [4, see references therein], the strategy facilitates in adapting the reliably decoded rate to the actual channel state without having any feedback link to the transmitter. In [13], the broadcasting approach is used in a system where there is no direct link between the source and the relay whereas in [14] the relay and the destination are collocated and cooperate over an AWGN channels with power constraints.

While the continuous infinite-layer transmission presents the upper bound on the achievable rate, [9] shows that 2-level superposition coding throughput approximates the infinite level superposition coding closely. A two-layer time division (TD) protocol for the fading relay channel is suggested by Yuksel and Erkip [8]. Katz and Shamai [1] consider a two-level transmission from the source to one of two collocated users where the user with the better channel to the source acts as a relay. Steiner and Shamai [19] use two-layer coding to derive achievable rates for a 2x2 MIMO.

By splitting the transmitted rate into two layer superposition coding and using the *sequential decode-and-forward (SDF)* [1],[3], several settings including the direct transmission, the MISO channel and the general relay channel are treated. Each layer is characterized by certain channel fading level and an allocated power. Two power allocation policies are examined, one with equal power distributions of the source and the relay and the other with separate power distributions of the terminals. For the MISO setting in the high SNR regime, the equal power allocation is shown to be optimal from diversity-multiplexing tradeoff point of view [15],[17]. The relay itself operates in either simplex or full-duplex modes. For the continuous broadcasting composed of infinite number of code layers, we derive lower bounds on the achievable rates based on the relay using a suboptimal power allocation. The 2x1 MISO as well as an oblivious source-relay cooperation are treated.

A few words regarding notation. Throughout the paper all logarithms are taken with respect to the base $e$. Rates are measured in [nats/ch. use]. The expectation operator is denoted by $E(\cdot)$.

The rest of the paper is organized as follows. Section II describes our considered model; Section III discusses the single layer SDF strategy for several scenarios; Section IV treats the continuous broadcasting lower bounds; Section V present the SDF strategy for a two-layer transmission and finally concluding remarks are given in Section VI. Proofs and derivations are deferred to Appendixes A-B.

## II. SYSTEM MODEL

The system consists of a source terminal $s$ who wishes to send information to a destination, denoted by $d$ (as depicted in Fig. 1). Due to the dynamic nature of the network, another terminal $r$ (which we refer to as relay) is occasionally present in close proximity to the source. However, the source is not aware whether or not the relay is actually present. It is assumed that whenever the relay is present, it is available to assist the transmission from the source to the destination. The information is transmitted over a shared wireless medium where transmissions received by the destination are subjected to block flat Rayleigh fading. The fading coefficients between the source and the destination and between the relay and the destination are denoted by $h_s$ and $h_r$, respectively, and are modeled by two independent zero mean unit variance complex Gaussian r.v's and are assumed constant over a coherence time equivalent of $N$ symbols, with $N$ large enough to allow Shannon theoretic arguments to hold, and are independent from one block to another.

Since the relay and the destination are physically collocated, the channel gain between the two is assumed to be $\sqrt{Q}e^{j\theta}$, where $Q$ is a fixed power gain (known to all) and $\theta$ is a random phase uniformly distributed over $[-\pi,\pi)$ also assumed fixed during each block of $N$ symbols and independent from one block to the next. Unless stated otherwise, all channel state information are assumed to be perfectly known by the respective receivers, and unknown to anyone else. Since the relay can cancel the phase $\theta$ without affecting the model statistics, we assume without loss of generality that $\theta = 0$.

During the transmission period of one fading block, the relay (if it exists) can assist the source in relaying the message to the destination. Unaware of relay's presence, the source assumes that in the worst case it is the only active transmitter, optimizing its transmission for the SISO channel, otherwise, there might be situations where the performance is worse than the optimal point-to-point signaling. When the relay exists, the received signals at the relay and the destination at time $n$, $n=1,2....N$, are modeled by

$$Y_r(n)=\sqrt{Q}X_s(n)+Z_r(n), Y_d(n)=h_s X_s(n)+h_r X_r(n)+Z_d(n). \qquad (1)$$

Otherwise, the signal at the destination is modeled by

$$Y_d(n)=h_s X_s(n)+Z_d(n). \qquad (2)$$

In (1) and (2) , $X_s(n)$ and $X_r(n)$ are the symbols transmitted during the $n$-th symbol interval by the source and the relay respectively, and $Z_r(n)$ and $Z_d(n)$ are the AWGN at the relay and the destination ,respectively, both modeled by i.i.d complex Gaussian r.v's with zero mean and unit variance. It will be convenient to denote the squared magnitude of the fading coefficients by $\nu_s=|h_s|^2$ and $\nu_r=|h_r|^2$ each of which is exponentially distributed with unit mean. Throughout the paper the source transmits with an average power $P_s$, and the relay (if present) transmits with an average power no larger than $P_r$, whenever it is active.

## III. SINGLE LAYER SEQUENTIAL DECODE-AND-FORWARD (SDF)

In the SDF strategy [1],[3] the source uses a single user capacity achieving codebook designed for the SISO channel when the relay is absent and transmits the codeword describing the message for the destination. The relay (if present) remains silent, while trying to decode the message. Once it is able to decode the message (after accumulating enough mutual information), it starts transmitting the same message using another predetermined codebook, acting as a second transmit antenna. If it is unable to decode the message before the block ends, it remains silent throughout the block, and no further cooperation takes place. The term sequential decode-and-forward is used to emphasize that the relay first decodes the entire message, and only then starts sending its codeword. Denote by $\varepsilon$ the fractional time within the transmission block when the relay is able to decode the message, i.e $\varepsilon \triangleq \min\left(1,\frac{R}{\log(1+P_s Q)}\right), \bar{\varepsilon}=1-\varepsilon$ ([1],[3]) . The general expected throughput is given in [3] by

$$R_{av}^{SDF}=R\cdot\begin{cases} e^{\frac{-e^R-1}{P_s}}+\int_0^{\frac{e^R-1}{P_s}} e^{-\frac{\left(e^{\frac{R-\varepsilon\log(1+\nu P_s)}{\bar{\varepsilon}}}-1-\nu P_s\right)}{P_r}} e^{-\nu}d\nu, R<\log(1+P_s Q) \\ e^{\frac{-e^R-1}{P_s}}, R>\log(1+P_s Q) \end{cases}. \qquad (3)$$

The optimal single user performance, given by [1] as $R_{av}^{SU}=R\cdot e^{\frac{e^R-1}{P_s}}$, will serve as a lower bound for the performance comparison. It is attained with $\varepsilon=1$ and the rate maximizing it is the attempted rate under oblivious cooperation given in [1]. A 2x1 MISO rate is attained for $\varepsilon = 0$. Two upper bounds on the SDF performance are the cut set upper bound [6] explicitly derived in [1] and the ergodic capacity of [14, Theorem 2].

## IV. Continuous broadcasting

The outage approach has a basic limitation of either allowing the attempted rate to be decoded or resulting in no information decoded in case of an outage. We now turn to examining an alternative approach known as continuous broadcasting. For completeness of presentation we give a quick overview of the method for a SISO channel and then present lower bounds on a relay channel under either a MISO or an oblivious cooperation scenarios.

### *A. Overview on single-user broadcasting*

Consider the SISO channel $y = h \cdot x + n$ where $y, x$ are the received and the transmitted signals respectively , $h$ is a fading coefficient known at the receiver only and $n$ is the additive noise. In the broadcast approach, the transmitter sends multilayer coded data matched to the equivalent fading parameter. The receiver decodes the maximal number of layers given a channel realization (per block). The differential rate per layer as a function of power allocation [4] is given by $dR(s) = \log\left(1 + \frac{s\rho(s)ds}{1+sI(s)}\right) = \frac{s\rho(s)ds}{1+sI(s)}$ where $I(s)$ is the residual interference function, such that $I(0) = P_s$. and $\rho(s) = -\frac{dI(s)}{ds}$ is the power allocation density function. Therefore, the maximal average rate over all fading realizations is

$$R_{bs} = \max_{I(u)} \int_0^\infty du \left(1 - F_s(u)\right) \frac{u\rho(u)}{1+uI(u)} \tag{4}$$

where $F_s(u)$ is the CDF of the equivalent fading gain random variable. It can be shown [4] that the optimal power allocation is given by

$$I_{opt}(u) = \begin{cases} P_s, u < u_0 \\ \frac{1 - F_s(u) - uf_s(u)}{u^2 f_s(u)}, u_0 < u < u_1 \\ 0, u > u_1 \end{cases} \tag{5}$$

where $u_0, u_1$ are obtained from the boundary conditions of $I_{opt}(u_0) = P_s$, $I_{opt}(u_1) = 0$.

### *B. Relay channel lower bounds*

Consider the problem of oblivious relaying where the transmitter performs continuous layering. It is assumed that when the relay exists, it first decodes the entire message from the source and then starts its transmission. With (1) as the channel model, in order to simplify the analysis we assume that the relay decoding time is negligible and hence this setting is called *informed SDF*. The informed SDF protocol assumes that the helping relay when available is informed of the transmitted packets in advance, and thus no time has to be spent on forwarding from source to relay. Thus when a relay is available it helps throughout the block.

Let us denote the power density at the transmitter by $\rho_s(s)$ and its corresponding residual interference function $I_s(s)$ where $I_s(s_0)=P_s$ and $I_s(s_1)=0$ (these are given already in closed form [4] ). The broadcasting power density at the relay by $\rho_r(s)$ and its residual interference function $I_r(s)$ are the subject for optimization. The relay power constraint is $I_r(s_0)=P_r$. We propose a suboptimal $I_r(s)$ of the form $I_r(s)=\frac{P_r}{P_s}I_s(s)$, which allows an analytical treatment of the problem. Under this power allocation, the equivalent fading parameter takes the form of $s \triangleq \nu_s + \frac{P_r}{P_s}\nu_r$. The PDF and the CDF of $s$ (1) are given by

$$\begin{array}{l} f(s)=\frac{e^{-\frac{s}{a}}}{a-1}+\frac{e^{-s}}{1-a}, F(s)=1+\frac{e^{-s}}{a-1}+\frac{ae^{-\frac{s}{a}}}{1-a}, a\triangleq\frac{P_r}{P_s}\neq 1, \\ f(s)=se^{-s}, F(s)=1-e^{-s}-se^{-s}, a=1 \end{array}.$$

Consider now an oblivious cooperation scenario when the source uses a SISO optimal power allocation [4] and the relay, if present, uses the mentioned $I_r(s)$. The only difference from the SISO here is that $F(s)$ is used for the rate evaluation. We call this setting *relay broadcasting*. Another option is a non-oblivious cooperation for which the source matches its power allocation to the equivalent fading parameter, i.e we have a 2x1 MISO with a suboptimal power allocations. This lower bound is termed as *MISO broadcasting*.

Fig. 2 displays the achievable rates display the achievable rates of the continuous broadcasting and the single layer transmission for various $\frac{P_r}{P_s}$ ratios. For the single layer it is assumed that when the relay is present , $\varepsilon=0$, corresponding to the MISO case in(3) . The stronger the relay, the greater is the rate loss due to sub-optimality of the relay layering power distribution $I_r(s)$.

## V. Two-layer SDF – Successive decoding

In the previous sections, we presented the achievable rates for the outage approach and for the continuous broadcasting. In the current section, we will incorporate the broadcast strategy into the SDF scheme by using two information streams only.

### A. General problem formulation

Consider $\eta_i, 1 \le i \le N$ to be the channel states for which the receiver is able to decode each of the $N$ information layers and let $\alpha_i, \beta_i \in [0,1], \sum_i \alpha_i = \sum_i \beta_i = 1$ be the power allocation parameters which govern the power allocation of the source and relay transmitters among the layers. Accordingly, we define the rates of the information streams to be

$$R_i = \log\left(1 + \frac{\eta_i \alpha_i P_s}{\eta_1 \bar{\alpha}_i P_s}\right), \bar{\alpha}_i = \sum_{k=i+1}^{N} \alpha_i, \eta_i < \eta_j \forall i < j \,. \qquad (6)$$

For $N = 2$, we extend the definitions of the relay decoding time $\varepsilon$ to the two layer situation by defining

$$R_1 = \log\left(\frac{1+\eta_1 P_s}{1+\eta_1 \bar{\alpha} P_s}\right), R_2 = \log\left(1+\eta_2 \bar{\alpha} P_s\right), \varepsilon_r^1 \triangleq \min\left(1, \frac{R_1}{\log\left(1+\frac{Q\alpha P_s}{Q\bar{\alpha} P_s}\right)}\right), \varepsilon_r^2 \triangleq \min\left(1, \max\left(\varepsilon_r^1, \frac{R_2}{\log\left(1+Q\bar{\alpha}P_s\right)}\right)\right). \quad (7)$$

The first term of (7), similarly to the single layer definition, involves the fractional time within the block where the relay gains sufficient mutual information to decode the first layer, and the maximization in the second term ensures that the second layer is decoded only after the first due to the required cancellation procedure. With the definition above, we define the mutual information at the destination for each of the layers. Consider the first layer, and let us assume that the relay transmits the first layer of its codeword with full power from the time it decodes this layer $\varepsilon_r^1$ to the time it decodes the second layer $\varepsilon_r^2$ and from there on it transmits according to the predetermined power allocation $\beta$. Therefore, for the first layer and second layers

$$\begin{aligned} I^{SDF,1} &= \varepsilon_r^1 \log\left(1+\frac{\nu_s \alpha P_s}{\nu_s \bar{\alpha} P_s}\right) + \left(\varepsilon_r^2 - \varepsilon_r^1\right)\log\left(1+\frac{\nu_s \alpha P_s + \nu_r P_r}{1+\nu_s \bar{\alpha} P_s}\right) + \left(1-\varepsilon_r^2\right)\log\left(1+\frac{\nu_s \alpha P_s + \nu_r \beta P_r}{1+\nu_s \bar{\alpha} P_s + \nu_r \bar{\beta} P_r}\right) \\ I^{SDF,2} &= \varepsilon_r^2 \log\left(1+\nu_s \bar{\alpha} P_s\right) + \left(1-\varepsilon_r^2\right)\log\left(1+\nu_s \bar{\alpha} P_s + \nu_r \bar{\beta} P_r\right) \end{aligned} \quad (8)$$

Note that for the second layer we have only the channel noise as the interference assuming successful decoding of the first layer. The overall average throughput can be computed by using (6)-(8) to obtain

$$R_{av}^{BSDF} = R_1 \cdot P\left[\left(I^{SDF,1} > R_1\right) \cap \left(I^{SDF,2} < R_2\right)\right] + \left(R_1 + R_2\right) \cdot P\left[\left(I^{SDF,1} > R_1\right) \cap \left(I^{SDF,2} > R_2\right)\right]. \quad (9)$$

Expression (9) can now be optimized over $\alpha, \beta, \eta_1, \eta_2$ for the given channel parameters to obtain the highest average throughput. We will assume unless specified otherwise that $\varepsilon_r^1 = \varepsilon_r^2$, implying simplex relay.

### B. Direct transmission

In the oblivious setting, the direct transmission rates are the ones used by the source and the average rate serves as the lower bound to achievable rates for the relay channel. The direct transmission is reflected by setting $\varepsilon_r^1 = \varepsilon_r^2 = 1$ in (8) ,thus reducing (9) to

$$R_{av}^{BSU} = R_1 P(\eta_1 < \nu_s < \eta_2) + (R_1 + R_2) P\left[(\nu_s > \eta_1) \cap (\nu_s > \eta_2)\right] = R_1 e^{-\eta_1} + R_2 e^{-\eta_2} \quad (10)$$

and similarly for the $N$ layer case $R_{av}^{BSU} = \sum_{i=1}^{N} R_i e^{-\eta_i}$ .

*C. Equal antenna layering power distribution MISO*

The MISO results are obtained by nullifying relay's decoding time, i.e $\varepsilon_r^1 = \varepsilon_r^2 = 0$ . In addition, we assume an equal power allocation $\alpha$ for the source and the relay. Under these, (9) becomes

$$R_{av}^{BMISO} = R_1 P\left[\left(\log\left(\frac{1+Y}{1+\bar{\alpha}Y}\right) > R_1\right) \cap \left(\log(1+\bar{\alpha}Y) < R_2\right)\right] + (R_1 + R_2) P\left[\left(\log\left(\frac{1+Y}{1+\bar{\alpha}Y}\right) > R_1\right) \cap \left(\log(1+\bar{\alpha}Y) > R_2\right)\right], Y \triangleq \nu_s P_s + \nu_r P_r \quad (11)$$

For the equal power allocation, the achievable rate can be computed via the distribution of $Y$ . The PDF of $Y$ is derived for two cases $P_s \neq P_r, P_s = P_r$ , resulting in

$$P(Y > u) = F_Y(u) = \begin{cases} \frac{1}{P_r - P_s}\left(P_r e^{-\frac{u}{P_r}} - P_s e^{-\frac{u}{P_s}}\right), P_s \neq P_r \\ \left(1 + \frac{u}{P_s}\right) e^{-\frac{u}{P_s}}, P_s = P_r \end{cases}, R_{av}^{BMISO} = R_1 F_Y(\eta_1 P_s) + R_2 F_Y(\eta_2 P_s) \quad (12)$$

and similarly for the $N$ layer case $R_{av}^{BMISO} = \sum_{i=1}^{N} R_i F_Y(\eta_i P_s)$ .

*D. Unequal antenna layering power distribution MISO*

In this section, a variable power allocation is employed where the relay is allowed to use an independent allocation parameter $\beta$ . The following result derived via explicit evaluation of the decoding probabilities quantifies the average throughput achievable by letting the relay use an independent power allocation.

*Theorem 1*: For a 2x1 MISO , a channel model described by (1) and independent predetermined power allocation coefficients $\alpha, \beta$ , the average throughput is given by

$$R_{av}^{BVMISO} = \begin{cases} R_1\left[e^{-\eta_1}\left(1 + \frac{1}{k-1}\right)\right] + R_2\left[\frac{e^{-\nu_{s,\alpha\beta 2} - k(\eta_1 - \nu_{s,\alpha\beta 2})}}{k-1} + e^{-\eta_2}\left(1 + \frac{1}{n-1}\right) - \frac{e^{-\nu_{s,\alpha\beta 2} - n(\eta_2 - \nu_{s,\alpha\beta 2})}}{n-1}\right], n \neq 1 \\ R_1\left[e^{-\eta_1}\left(1 + \frac{1}{k-1}\right)\right] + R_2\left[\frac{e^{-\nu_{s,\alpha\beta 2} - k(\eta_1 - \nu_{s,\alpha\beta 2})}}{k-1} + e^{-\eta_2}\left(1 + \eta_2 - \nu_{s,\alpha\beta 2}\right)\right], n = 1 \end{cases}, 1 - e^{R_1}\bar{\beta} < 0$$

$$, n \triangleq \frac{\bar{\alpha} P_s}{\bar{\beta} P_r}, k \triangleq \frac{\alpha P_s}{\left(\beta + \eta_1 P_s(\beta - \alpha)\right) P_r}$$

$$\nu_{s,\alpha\beta 1} \in [0, \eta_1] \triangleq \begin{cases} 0, \frac{\bar{\alpha}\eta_2}{\bar{\beta}} > \frac{\alpha\eta_1}{\beta + \eta_1 P_s(\beta - \alpha)} \\ -\left(\frac{\alpha\eta_1\bar{\beta} - \bar{\alpha}\eta_2\left(\beta + \eta_1 P_s(\beta - \alpha)\right)}{\bar{\alpha}\left(\beta + \eta_1 P_s(\beta - \alpha)\right) - \alpha\bar{\beta}}\right), else \end{cases}, \nu_{s,\alpha\beta 2} \in [\eta_1, \eta_2] \triangleq -\left(\frac{\alpha\eta_1\bar{\beta} - \bar{\alpha}\eta_2\left(\beta + \eta_1 P_s(\beta - \alpha)\right)}{\bar{\alpha}\left(\beta + \eta_1 P_s(\beta - \alpha)\right) - \alpha\bar{\beta}}\right)$$

$$R_{av}^{BVMISO}=\begin{cases} R_1\left[e^{-\eta_1}\left(1+\frac{1}{k-1}\right)-\frac{e^{-k\eta_1}}{k-1}\right]+R_2\left[\frac{e^{-\nu_{s,\alpha\beta 1}-k\left(\eta_1-\nu_{s,\alpha\beta 1}\right)}}{k-1}-\frac{e^{-k\eta_1}}{k-1}+e^{-\eta_2}\left(1+\frac{1}{n-1}\right)-\frac{e^{-\nu_{s,\alpha\beta 1}-n\left(\eta_2-\nu_{s,\alpha\beta 1}\right)}}{n-1}\right],n,k\neq 1 \\ R_1\left[e^{-\eta_1}\left(1+\frac{1}{k-1}\right)-\frac{e^{-k\eta_1}}{k-1}\right]+R_2\left[\frac{e^{-\nu_{s,\alpha\beta 1}-k\left(\eta_1-\nu_{s,\alpha\beta 1}\right)}}{k-1}-\frac{e^{-k\eta_1}}{k-1}+e^{-\eta_2}\left(1+\eta_2-\nu_{s,\alpha\beta 1}\right)\right],n=1,k\neq 1 \\ R_1\left[e^{-\eta_1}\left(1+\eta_1\right)\right]+R_2\left[e^{-\eta_1}\nu_{s,\alpha\beta 1}+e^{-\eta_2}\left(1+\frac{1}{n-1}\right)-\frac{e^{-\nu_{s,\alpha\beta 1}-n\left(\eta_2-\nu_{s,\alpha\beta 1}\right)}}{n-1}\right],n\neq 1,k=1 \\ R_1\left[e^{-\eta_1}\left(1+\eta_1\right)\right]+R_2\left[e^{-\eta_1}\nu_{s,\alpha\beta 1}+e^{-\eta_2}\left(1+\eta_2-\nu_{s,\alpha\beta 1}\right)\right],n=k=1 \end{cases} \quad ,1-e^{R_1}\overline{\beta}>0 \tag{13}$$

By Theorem.1, it is evident that the relay's power allocation has a crucial effect on the achievable rate, and a powerful relay does not guarantee a high achievable rate unless equipped with an appropriate power allocation. For an equal power allocation (13) reduces to (12) as $1-e^{R_1}\overline{\alpha}>0$ by definition and $\nu_{s,\alpha\beta 1}=0$.Examination of the achievable rates leads to the following theorem.

*Theorem 2* : For a 2x1 MISO , a channel model described by (1) and independent predetermined power allocation coefficients $\alpha,\beta$, if $n,k\geq 1$ the maximal achievable average throughput is

$$R_{av}^{\max MISO}=R_1\left[e^{-\eta_1}\left(1+\eta_1\right)\right]+R_2\left[e^{-\eta_1}\nu_{s,\alpha\beta 1}+e^{-\eta_2}\left(1+\eta_2-\nu_{s,\alpha\beta 1}\right)\right] \tag{14}$$

achieved for $n=k=1$. If $n,k<1$ a maximal throughput of $R_1+R_2$ is achieved with $n=k=0$.

*Proof:* The proof is trivial by using (15) and is omitted.

$$\begin{aligned} &e^{-\eta_1}\left(1+\frac{1}{k-1}\right)-\frac{e^{-k\eta_1}}{k-1}\leq e^{-\eta_1}\left(1+\eta_1\right) \\ &\frac{e^{-\nu_{s,\alpha\beta 1}-k\left(\eta_1-\nu_{s,\alpha\beta 1}\right)}}{k-1}-\frac{e^{-k\eta_1}}{k-1}+e^{-\eta_2}\left(1+\frac{1}{n-1}\right)-\frac{e^{-\nu_{s,\alpha\beta 1}-n\left(\eta_2-\nu_{s,\alpha\beta 1}\right)}}{n-1}\leq e^{-\eta_1}\nu_{s,\alpha\beta 1}+e^{-\eta_2}\left(1+\eta_2-\nu_{s,\alpha\beta 1}\right) \end{aligned} \quad . \tag{15}$$

The average throughput can not be larger than $R_1+R_2$ which is achieved for $n=k=0$, thus resulting in a maximal throughput overall□ .

Theorem 2 leads to a lemma regarding the optimal power allocation.

*Lemma 1*: For $P_s \ll P_r$ the optimal power allocation is $\alpha=\beta$.

*Proof* : Set $\alpha=\beta$. By (13) , $n=k=\frac{P_s}{P_r}\to 0$, resulting in a maximal achievable rate and therefore optimal.

A step in the direction of determining an optimal power allocation for the MISO is taken in the following theorem proven in Appendix A where we establish the optimal power allocation for an asymptotic source power and a constant ratio of source to relay powers.

*Theorem 3* : For a 2x1 MISO setting satisfying $P_s\to\infty,\frac{P_s}{P_r}=c$ under the channel model described by (1) , the equal power allocation is optimal.

*E. Simplex relay – equal antenna layering power distribution*

In this section, we assume that $0<\varepsilon_r^1=\varepsilon_r^2<1$ and derive the corresponding achievable rates. We start with the equal power allocation for which there are no degrees of freedom to optimize under the oblivious setting as the rates and $\alpha$ are completely dictated by (10). Starting with the first layer, applying $\alpha=\beta$ , $\varepsilon_r^1=\varepsilon_r^2$ to (8) , and rearranging, we arrive to

$$I^{SDF,1}>R_1 \Rightarrow \varepsilon_r^2 < \underbrace{\frac{\log\left(1+\frac{\nu_s\alpha P_s+\nu_r\alpha P_r}{1+\nu_s\bar{\alpha}P_s+\nu_r\bar{\alpha}P_r}\right)-R_1}{\log\left(1+\frac{\nu_s\alpha P_s+\nu_r\alpha P_r}{1+\nu_s\bar{\alpha}P_s+\nu_r\bar{\alpha}P_r}\right)-\log\left(1+\frac{\nu_s\alpha P_s}{1+\nu_s\bar{\alpha}P_s}\right)}}_{F_1(\cdot)}, \nu_s<\eta_1, \qquad (16)$$

By conditioning on the value of $\nu_s$, the previous expression is equivalent to

$$P\left[\left(\varepsilon_r^1 \le F_1\left(\nu_s,\nu_r,\alpha,P_s,P_r,R_1\right)\right)\cap\left(\left(\frac{R_2}{\log\left(1+Q\bar{\alpha}P_s\right)} \le F_1\left(\nu_s,\nu_r,\alpha,P_s,P_r,R_1\right)\right)\right)\right].(17)$$

Bearing in mind that the decoding of the second layer is possible only in the case of successfully decoding the first layer priory, we arrive from (8) to

$$I^{SDF,2}>R_2 \Rightarrow \varepsilon_r^2 < \frac{\log\left(1+\nu_s\bar{\alpha}P_s+\nu_r\bar{\alpha}P_r\right)-R_2}{\log\left(1+\nu_s\bar{\alpha}P_s+\nu_r\bar{\alpha}P_r\right)-\log\left(1+\nu_s\bar{\alpha}P_s\right)} \qquad (18)$$

and the outage probability is zero for $\nu_s>\eta_2$ as $\log\left(1+\nu_s\bar{\alpha}P_s\right)>R_2$ and even the single user would decode both layers under those conditions. By defining $M\triangleq\nu_r\bar{\alpha}P_r, Z\triangleq 1+\nu_s\bar{\alpha}P_s$, (18) becomes

$$P\left[\left(\nu_r>\frac{Z\left[\left(Ze^{-R_2}\right)^{\frac{1}{X_1-1}}-1\right]}{\bar{\alpha}P_r}\right)\cap\left(\nu_r>\frac{Z\left[\left(Ze^{-R_2}\right)^{\frac{1}{X_2-1}}-1\right]}{\bar{\alpha}P_r}\right)\right], \frac{Z\left[\left(Ze^{-R_2}\right)^{\frac{1}{X-1}}-1\right]}{\bar{\alpha}P_r} \triangleq U\left(\nu_s,P_s,P_r,\alpha,\bar{\alpha},R_2,X\right), X=\left(\underbrace{\varepsilon_r^1}_{X_1},\underbrace{\min\left(1,\frac{R_2}{\log\left(1+Q\bar{\alpha}P_s\right)}\right)}_{X_2}\right)$$

(19)

The function $U$ is an increasing function of $X$ for $\left(X<1, Ze^{-R_2}<1\right)$(Appendix B), leading to the intersection given by $U\left(\cdot,..\varepsilon_r^2\right)$. Returning to the first layer, we define $L\triangleq\nu_rP_r, S\triangleq\nu_sP_s$ and rewrite each of the terms of (17) to the equivalent form of

$$\left\{\nu_r \gtrless \frac{1}{P_r}\left[-S-\left(\frac{1-t}{1-t\bar{\alpha}}\right)\right] \triangleq F\left(\nu_s,\alpha,P_s,P_r,X,R_1\right) \gtrless 0, 1-t\bar{\alpha}\gtrless 0, t\triangleq\left[\left(e^{-R_1}\left(\frac{1+S}{1+\bar{\alpha}S}\right)\right)^{\frac{1}{X-1}}\left(\frac{1+S}{1+\bar{\alpha}S}\right)\right]. \qquad (20)\right.$$

From Lemma 2 of Appendix B , intersection (20) evaluates to

$$P\left(\nu_r>\frac{1}{P_r}\left(-S-\left(\frac{1-t_{\max(X_1,X_2)=\varepsilon_r^2}}{1-t_{\max(X_1,X_2)=\varepsilon_r^2}\bar{\alpha}}\right)\right)\middle|\nu_{s,dc}<\nu_s<\eta_1\right). \qquad (21)$$

The decoding probability of the first layer only as well as for both layers can be computed explicitly by using (19)-(21). The exact form of the solution involving the intersection terms is dependent upon the behavior of the $U(\cdot),F(\cdot)$ functions over the $[0,\eta_1]$ interval, as explained in Appendix B. For the achievable rate we have the following theorem proven in Appendix B.

*Theorem 4* : For a relay channel described by (1) and equal antenna layering power allocation distribution coefficient $\alpha$, the average throughput is given by

$$\begin{aligned}
R_{av}^{RB} &= R_1\left(e^{-\eta_1} + \int_{\nu_{s,dc}}^{\eta_1} e^{-F-\nu_s} d\nu_s\right) + R_2\left(\int_{\eta_1}^{\eta_2} e^{-U-\nu_s} d\nu_s + e^{-\eta_2}\right) + R_2\left(\sum_{k=0}^{\frac{n}{2}} \int_{\nu_{s,2k}}^{\nu_{s,2k+1}} e^{-U-\nu_s} d\nu_s + \sum_{k=0}^{\frac{n}{2}-1} \int_{\nu_{s,2k+1}}^{\nu_{s,2(k+1)}} e^{-F-\nu_s} d\nu_s\right) \\
R_{av}^{RB} &= R_1\left(e^{-\eta_1} + \int_{\nu_{s,dc}}^{\eta_1} e^{-F-\nu_s} d\nu_s\right) + R_2\left(\int_{\eta_1}^{\eta_2} e^{-U-\nu_s} d\nu_s + e^{-\eta_2}\right) + R_2\left(\sum_{k=0}^{\frac{m-1}{2}} \int_{\nu_{s,2k}}^{\nu_{s,2k+1}} e^{-F-\nu_s} d\nu_s + \sum_{k=0}^{\frac{m-1}{2}} \int_{\nu_{s,2k+1}}^{\nu_{s,2(k+1)}} e^{-U-\nu_s} d\nu_s\right) \qquad (22) \\
& F(\nu_{s,k}\ldots) = U(\nu_{s,k}\ldots), k = 0,1..n(m) \\
& \nu_{s,0} = \nu_{s,dc}, \nu_{s,n+1} = \nu_{s,m+1} = \eta_1
\end{aligned}$$

where $n$ is even, $m$ is odd and they quantify the number of intersection points between $F(\cdot),U(\cdot)$ over the $[0,\eta_1]$ interval.

Note that for $Q\to\infty$, (22) will not necessary approach the MISO rate as the source-relay channel capacity for the first layer is interference limited by the second layer so the decoding times will be finite.

### *F. Simplex relay – unequal antenna layering power distribution*

Next, we evaluate (9), where the unequal layering power distribution introduces a degree of freedom in the form of the $\beta$ allocation coefficient to be optimized. With $\beta$ introduced, we rearrange the general expression for the second layer (8) to obtain

$$I^{SDF,2} > R_2 \Rightarrow \varepsilon_r^2 < \frac{\log\left(1+\nu_s\bar{\alpha}P_s+\nu_r\bar{\beta}P_r\right)-R_2}{\log\left(1+\nu_s\bar{\alpha}P_s+\nu_r\bar{\beta}P_r\right)-\log\left(1+\nu_s\bar{\alpha}P_s\right)}, \quad (23)$$

and by defining this time $M \triangleq \nu_r\bar{\beta}P_r, Z \triangleq 1+\nu_s\bar{\alpha}P_s$, (19) remain valid with the appropriate substitution of $\bar{\beta}$ for $\bar{\alpha}$ in the denominator of $U\left(\nu_s,P_s,P_r,\alpha,\bar{\beta},R_2,X\right)$. The outage probability remains zero for $\nu_s>\eta_2$. For the first layer , the non-outage expression becomes

$$I^{SDF,1} > R_1 \Rightarrow \varepsilon_r^2 \lesseqgtr \frac{\log\left(1+\frac{\nu_s\alpha P_s+\nu_r\beta P_r}{1+\nu_s\bar{\alpha}P_s+\nu_r\bar{\beta}P_r}\right)-R_1}{\log\left(1+\frac{\nu_s\alpha P_s+\nu_r\beta P_r}{1+\nu_s\bar{\alpha}P_s+\nu_r\bar{\beta}P_r}\right)-\log\left(1+\frac{\nu_s\alpha P_s}{1+\nu_s\bar{\alpha}P_s}\right)}, s.t \log\left(1+\frac{\nu_s\alpha P_s+\nu_r\beta P_r}{1+\nu_s\bar{\alpha}P_s+\nu_r\bar{\beta}P_r}\right) \gtrless \log\left(1+\frac{\nu_s\alpha P_s}{1+\nu_s\bar{\alpha}P_s}\right). \quad (24)$$

The last term of (24) is equivalent to $\beta \gtrless \nu_s P_s(\alpha-\beta)$. Denote $\nu_s^* \triangleq \frac{\beta}{P_s(\alpha-\beta)}$, then

$$\begin{cases}\beta \lesseqgtr \nu_s P_s\left(\alpha-\beta\right), \nu_s \gtrless \nu_s^*, \nu_s^*>0 \\ \beta \gtreqless \nu_s P_s\left(\alpha-\beta\right), \nu_s \gtrless \nu_s^*, \nu_s^*<0\end{cases}.$$

Examination of (24) reveals that taking $\beta<\alpha$ may result in low decoding probabilities irrespective of relay's power, as the additional interference of the second layer to the first layer originated by relay's transmission causes an overall degradation. High $P_r$ will contribute more interference when $\beta<\alpha$. Obviously, one could use such $\beta$ and compute the achievable rates, but as we look for the optimal power allocation, we will assume $\beta \geq \alpha$. Keeping the definitions used for the equal power allocation, we arrive to the general form of $\nu_r \gtrless \frac{1}{P_r}\left(-S\left(\frac{1-t\bar{\alpha}}{1-t\bar{\beta}}\right)-\left(\frac{1-t}{1-t\bar{\beta}}\right)\right) \triangleq K\left(\nu_s,\alpha,\beta,P_s,P_r,X,R_1\right), 1-t\bar{\beta} \gtrless 0$. Using the results and definitions of Appendix B with the constraint $\beta \geq \alpha$ and following along the derivation lines of the previous section with the relevant probability expressions, we have the following theorem (Appendix B).

*Theorem 5* : For a relay channel described by (1) and unequal antenna layering power allocation distribution coefficients $\alpha,\beta$ such that $\beta \geq \alpha$, the average throughput is given by Theorem 4 with the substitutions $F(\cdot) \triangleq K(\cdot), \nu_{s,dc} \triangleq \nu_{s,dcv}$

## G. *Full-Duplex relay – unequal antenna layering power distribution*

A full-duplex relay transmits the first layer of it's codeword as soon as it has decoded the first layer of source's transmission with full available power. After decoding the second layer, the power is split among the layers according to $\beta$. As the second layer mutual information is not affected by the full-duplex action, (23) remains relevant and we concentrate on the evaluation of the first layer by rewriting (8) as

$$\varepsilon_r^2 \leq \frac{\log\left(\frac{1+S+L}{1+\bar{\alpha}S+\bar{\beta}L}\right)-R_1+\varepsilon_r^1\left[\log\left(\frac{1+S}{1+\bar{\alpha}S}\right)-\log\left(\frac{1+S+L}{1+\bar{\alpha}S}\right)\right]}{\left[\log\left(\frac{1+S+L}{1+\bar{\alpha}S+\bar{\beta}L}\right)-\log\left(\frac{1+S+L}{1+\bar{\alpha}S}\right)\right]}, L \triangleq \nu_r P_r, S \triangleq \nu_s P_s \quad (25)$$

and by using a now standard procedure of evaluation for $\varepsilon_r^2$ with the definitions for $X_1, X_2$, one gets by plugging $X_2$ into (25) and defining $\xi \triangleq X_2 - X_1$

$$\nu_r \gtrless \frac{1}{P_r}\left(-S\left(\frac{1-t_{X_1} f(L)\bar{\alpha}}{1-t_{X_1} f(L)\bar{\beta}}\right)-\left(\frac{1-t_{X_1}}{1-t_{X_1} f(L)\bar{\beta}}\right)\right), 1-t_{X_1} f(L)\bar{\beta} \gtrless 0, f(L) \triangleq \left(1+\frac{\bar{\beta}L}{1+\bar{\alpha}S}\right)^{\frac{\xi}{X_1-1}} \quad (26)$$

Due to the nonlinear dependency of the RHS in (26) on $L$, we can no longer get an expression dependent solemnly on $\nu_s$ over which we condition. Therefore, the results presented in the next section are obtained by direct evaluation of the rate expression (9) via Monte-Carlo technique. The main result will show that unless the source-relay channel gain as well as source power are low, no gain is obtained by the full-duplex relay over the simplex as

$$\frac{R_{1,opt}}{\log\left(1+\frac{Q\alpha_{opt}P_s}{1+Q\bar{\alpha}_{opt}P_s}\right)} > \frac{R_{2,opt}}{\log\left(1+Q\bar{\alpha}_{opt}P_s\right)} \qquad (27)$$

for the optimal single user rates. Since those rates are found numerically, a general analytical proof of (27) is not available. From the examination of the optimal single user rates , $R_{1,opt} > R_{2,opt}$, but proving this analytically remains an open problem and can be stated as a conjecture. Based on the conjecture, a sufficient condition for (27) to hold is

$$1 < 2\bar{\alpha}_{opt} + Q\bar{\alpha}_{opt}^2 P_s . \qquad (28)$$

*H. Numerical results*

In this section we present the numerical results for the broadcasting SDF. In Fig. 3 the two layered transmission closes about 80% of the gap to the continuous upper bound. Fig. 4 displays the rates for the 2x1 MISO under both equal and unequal power distributions. Note that for small source to relay power ratios the equal power allocation emerges as optimal. Two-layered transmission again bridges most of the gap to the continuous lower bound, which is shown to be tightly approximated by $N = 8$ layers, suggesting that in fact the sub-optimal allocation might be optimal under those channel conditions. Depicted in Fig. 5 is a channel condition when the unequal distribution results in very substantial gains over its equal counterpart. This can be partially explained by noting that for a high source to relay power ration the equal power allocation will lead to single user rates which are sub-optimal. Fig. 6 presents the SDF two layer rates for various source-relay channel qualities. Since the direct transmission rate does not depend on the relay power and the source-relay channel gain, the influence of $Q$ is expected to be minor. The achieved gain is about 2dB for low $P_s$,and decreases with the rate. The gain decrease can be explained by the increase of the optimal attempted rates for the direct transmission with $P_s$. This, in turn, increases the decoding times, as the capacity of the source-relay channel capacity saturates faster than the rate growth. As a result, the portion of the block in which the relay contributes decreases, eventually making the relay channel appear like SISO. Fig. 7 demonstrates the achievable rates when $P_s$ is fixed and $P_r$ is varied, for two $P_s$ values of 10dB and 20dB. It can be seen that relay's effect on the rate depends on $\frac{P_r}{P_s}$ much stronger than on it's absolute power. For the weaker source, the curves exhibit a saturation behavior as the relay power increases. This is since the rates are dictated by the SISO channel and the overall rate can not be larger than the attempted $(R_1 + R_2)$, therefore saturation will occur at some point. Fig. 8 presents the SDF two layer rates for variable power allocation Note that by comparison with Fig. 6 , the gain decreases slower, probably due to a more significant contribution by the relay achieved via the optimization. The gain over the equal power allocation is about 0.4dB. As mentioned the 2x1 MISO bound can not be attained as taking $Q \to \infty$ does not result in zero decoding times by (7). Finally, the full duplex results of Fig. 9 show that the gain from this mode is limited to low collocation gains and as we assume a tight coupling of the source and the relay, the simplex mode is sufficient for optimal oblivious cooperation.

## VI. CONCLUSION

We have demonstrated the incorporation of a two-layered broadcasting strategy into the SDF oblivious protocol for the relay channel. By using two layers of information, substantial gains were shown to be achievable over the outage approach for various system settings. Optimality of the equal power allocation as well as a relay simplex mode of functioning was demonstrated under several channel conditions, simplifying design matters. Continuous broadcasting rates were shown to be high even under sub-optimal power allocation by the relay, as well as tightly approximated by a discrete transmission with limited number of layers. In the second part of the paper we will use the Block-Markov scheme to overcome some of the SDF limitations.

## APPENDIX A

## PROOF OF THEOREM 3

We examine the MISO channel in the $SNR \gg 1$ regime, by employing the notions of the diversity gain and multiplexing gain [15]. A coding scheme $C_{SNR}$ is said to achieve a *spatial multiplexing gain* $r$ and *multiplexing diversity* $d$ if the data rate and the error probability satisfy [15]

$$\lim_{SNR\to\infty} \frac{R(SNR)}{\log(SNR)} = r,\ \lim_{SNR\to\infty} \frac{\log(P_e(SNR))}{\log(SNR)} = -d \qquad (29)$$

The result of [15, Theorem 4] quantifies the outage probability of the multi-antenna channel at certain rate for a $m \times n$ multiple antenna channel described by

$$Y = HX + N \qquad (30)$$

and a transmission rate of $R = r\log(SNR), r < \min(m,n)$. The computation is performed via the distribution of the ordered eigenvalues of $HH^{\dagger}$. In our setting $m = 2, n = 1$ hence $r \le 1$. In addition , $HH^{\dagger} = \lambda_{1x1} = |h_s|^2 + |h_d|^2$. The PDF of $\lambda$ is $\Gamma(2,1) = xe^{-x}$ and the CDF is $1 - e^{-x} - xe^{-x}$ , approximated for small $x$ as $P(\lambda < x) = x^2, x \ll 1$. By [15, Theorem 2], the optimal tradeoff curve for the 2x1 system satisfies $d + 2r = 2, r < 1$ and it will serve as an upper bound. As for $P_s \to \infty$ we want the first layer to capture all the power up to a fracture allocated to the second as to make the scheme interference free, we use an extension of an allocation presented in [17] for the SISO channel and allocate the layers' power according to

$$P_s(1+c)-P_s^{\alpha}-(cP_s)^{\beta},\, P_s^{\alpha}+(cP_s)^{\beta} \qquad (31)$$

where $\alpha \neq \beta$ in general. Following [25],[30], for the first layer we get

$$P_{out}(r_1\log(SNR)) = P\left(\frac{1+P_s(1+c)\lambda}{1+\left(P_s^{\alpha}+(cP_s)^{\beta}\right)\lambda} < \left(P_s(1+c)\right)^{r_1}\right) \approx P\left(P_s(1+c)\lambda < \left(P_s(1+c)\right)^{r_1}\left(1+\left(P_s^{\alpha}+(cP_s)^{\beta}\right)\lambda\right)\right)$$

$$= P\left(\lambda < \frac{\left(P_s(1+c)\right)^{-(1-r_1)}}{1-\left(P_s(1+c)\right)^{-(1-r_1)}\left(P_s^{\alpha}+(cP_s)^{\beta}\right)}\right), 1-\left(P_s(1+c)\right)^{-(1-r_1)}\left(P_s^{\alpha}+(cP_s)^{\beta}\right)>0,\; P\left(\lambda > \frac{\left(P_s(1+c)\right)^{-(1-r_1)}}{1-\left(P_s(1+c)\right)^{-(1-r_1)}\left(P_s^{\alpha}+(cP_s)^{\beta}\right)}\right)=1, else$$

Due to $P_s \to \infty$, the dominant term in the denominator will be the one with the larger exponent. We therefore examine three possible cases, two of which are symmetric: $\alpha=\beta, \alpha>\beta, \beta<\alpha$. Starting with $\alpha=\beta$, the last expression becomes

$$P\left(\lambda < \frac{\left(P_s(1+c)\right)^{-(1-r_1)}}{1-\left(P_s(1+c)\right)^{-(1-r_1)}\left(P_s^{\alpha}+(cP_s)^{\beta}\right)}\right) = P\left(\lambda < \frac{\left(P_s(1+c)\right)^{-(1-r_1)}}{1-\left(P_s(1+c)\right)^{-(1-r_1)}\left(P_s^{\alpha}\left(1+c^{\alpha}\right)\right)}\right) = P\left(\lambda < \frac{P_s^{-(1-r_1)}(1+c)^{-(1-r_1)}}{1-P_s^{-(1-r_1-\alpha)}(1+c)^{-(1-r_1)}\left(1+c^{\alpha}\right)}\right) \qquad (32)$$

For $(1-r_1-\alpha)<0$, (32) evaluates to

$$P\left(\lambda > -\frac{1}{\left(1+c^{\alpha}\right)}P_s^{-\alpha}\right)=1 \qquad (33)$$

and for $(1-r_1-\alpha)>0$ to

$$P\left(\lambda < (1+c)^{-(1-r_1)}P_s^{-(1-r_1)}\right) = (1+c)^{-2(1-r_1)}P_s^{-2(1-r_1)}. \qquad (34)$$

For $\alpha>\beta$, the outage probability is (34) for $(1-r_1-\alpha)>0$ and for $(1-r_1-\alpha)<0$ we get

$$P\left(\lambda > -P_s^{-\alpha}\right)=1. \qquad (35)$$

Similarly, for $\beta>\alpha$, the outage probability for $(1-r_1-\beta)>0$ is (34) and for $(1-r_1-\beta)<0$

$$P\left(\lambda > -\frac{1}{c^{\beta}}P_s^{-\beta}\right)=1. \qquad (36)$$

Turning now to the evaluation of the outage probability for the second layer,

$$P_{out}(r_2\log(SNR)) = P\left(1+\left(P_s^{\alpha}+(cP_s)^{\beta}\right)\lambda < \left(P_s(1+c)\right)^{r_2}\right) \approx P\left(\lambda < \frac{\left((1+c)P_s\right)^{r_2}}{\left(P_s^{\alpha}+(cP_s)^{\beta}\right)}\right). \qquad (37)$$

For $\alpha=\beta$, (37) becomes

$$\begin{cases} 1, \alpha < r_2 \\ \frac{(1+c)^{2r_2}}{(1+c^{\alpha})^2} P_s^{-2(\alpha-r_2)} = \frac{(1+c)^{2r_2}}{(1+c^{\beta})^2} P_s^{-2(\beta-r_2)}, \alpha > r_2 \end{cases}, \tag{38}$$

for $\alpha > \beta$ we have

$$\begin{cases} 1, \alpha < r_2 \\ (1+c)^{2r_2} P_s^{-2(\alpha-r_2)}, \alpha > r_2 \end{cases} \tag{39}$$

since $P_s^{\alpha}$ is the dominant term, and finally for $\beta > \alpha$ the dominance of the $(cP_s)^{\beta}$ term leads to

$$\begin{cases} 1, \beta < r_2 \\ \frac{1}{c^{2\beta}} (1+c)^{2r_2} P_s^{-2(\beta-r_2)}, \beta > r_2 \end{cases} . \tag{40}$$

Let us now evaluate the achievable average rates under each of the allocations by computing

$$R_{av} = (1 - P_{out}(r_1)) r_1 + (1 - P_{out}(r_2)) r_2$$

$$\alpha = \beta : R_{av} = \begin{cases} 1 - r_1 - \alpha < 0, \alpha < r_2 \rightarrow R_{av} = 0 \\ 1 - r_1 - \alpha < 0, \alpha > r_2 \rightarrow R_{av} = 0 \\ 1 - r_1 - \alpha > 0, \alpha < r_2 \rightarrow \left(1 - (1+c)^{-2(1-r_1)} P_s^{-2(1-r_1)}\right) r_1 \\ 1 - r_1 - \alpha > 0, \alpha > r_2 \rightarrow \left(1 - (1+c)^{-2(1-r_1)} P_s^{-2(1-r_1)}\right) r_1 + \left(1 - \frac{(1+c)^{2r_2}}{(1+c^{\alpha})^2} P_s^{-2(\alpha-r_2)}\right) r_2 \end{cases}$$

$$\alpha > \beta : R_{av} = \begin{cases} 1 - r_1 - \alpha < 0, \alpha < r_2 \rightarrow R_{av} = 0 \\ 1 - r_1 - \alpha < 0, \alpha > r_2 \rightarrow R_{av} = 0 \\ 1 - r_1 - \alpha > 0, \alpha < r_2 \rightarrow \left(1 - (1+c)^{-2(1-r_1)} P_s^{-2(1-r_1)}\right) r_1 \\ 1 - r_1 - \alpha > 0, \alpha > r_2 \rightarrow \left(1 - (1+c)^{-2(1-r_1)} P_s^{-2(1-r_1)}\right) r_1 + \left(1 - (1+c)^{2r_2} P_s^{-2(\alpha-r_2)}\right) r_2 \end{cases}$$

$$\alpha < \beta : R_{av} = \begin{cases} 1 - r_1 - \beta < 0, \beta < r_2 \rightarrow R_{av} = 0 \\ 1 - r_1 - \beta < 0, \beta > r_2 \rightarrow R_{av} = 0 \\ 1 - r_1 - \beta > 0, \beta < r_2 \rightarrow \left(1 - (1+c)^{-2(1-r_1)} P_s^{-2(1-r_1)}\right) r_1 \\ 1 - r_1 - \beta > 0, \beta > r_2 \rightarrow \left(1 - (1+c)^{-2(1-r_1)} P_s^{-2(1-r_1)}\right) r_1 + \left(1 - \frac{1}{c^{2\beta}} (1+c)^{2r_2} P_s^{-2(\beta-r_2)}\right) r_2 \end{cases} \tag{41}$$

From (38)-(40) we see that for a multiplexing gain higher than the power allocation coefficient for the first layer we can not decode the second layer at all. In addition, $r_2$ can be decoded only if $r_1$ can be decoded and therefore an average rate of zero is achieved if $r_1$ is un-decodable. Examination of (41) reveals that the equal power allocation is optimal. The relevant expressions are those with $(\beta, \alpha) > r_2, (1-\beta, 1-\alpha) > r_1$ and for high SNR the optimal coefficients are arbitrary close to 1 to express the fact that the allocation of the first layer must capture nearly all power available. Comparing the achievable diversity-multiplexing curve to the optimal curve, for the two level coding we have

$$d_1 + 2r_1 = 2, r_1 < 1 - \alpha, d_2 + 2r_2 = 2\alpha, r_2 < \alpha \tag{42}$$

meaning that for an arbitrary small $\varepsilon > 0$ and $\alpha = 1-\varepsilon$ one can achieve multiplexing gains as much as $r_2 = 1-\varepsilon$ and simultaneously high diversity $d_1 = 2(1-\varepsilon)$.

## APPENDIX B

### PROOFS OF THEOREMS 4,5.

From (20) $\frac{\partial t}{\partial \nu_s} = \left( e^{\frac{R_1}{1-X}} \left( \frac{1+S}{1+\bar{\alpha}S} \right)^{\frac{X}{X-1}} \right) \left( \frac{\alpha P_s}{(1+S)(1+\bar{\alpha}S)} \right) \frac{X}{X-1} < 0$ so $t$ is monotonically decreasing w.r.t $\nu_s$. This means that $t \geq e^{R_1}, \nu_s < \eta_1$ as $t(\eta_1,..) = e^{R_1}$. Next, we define $f(t) \triangleq -\left( \frac{1-t}{1-t\bar{\alpha}} \right)$, which is also a monotonically decreasing function of $\nu_s$ with a possible second kind discontinuity. To show this, $\frac{\partial f}{\partial \nu_s} = \frac{\partial f}{\partial t}\frac{\partial t}{\partial \nu_s} = \frac{\alpha}{(1-t\bar{\alpha})^2}\frac{\partial t}{\partial \nu_s} < 0$ and the differential is discontinuous at $t = \frac{1}{\bar{\alpha}}$. The behavior of the functional $F(\nu_s, \alpha, P_s, P_r, X, R_1)$ (20) dictates the bounds for the probability (17) to be positive and separates the range of $[0, \eta_1]$ into disjoint sets. Note that $\frac{\partial F}{\partial t} = \frac{\alpha}{(1-t\bar{\alpha})^2} \geq 0$ so $F(\cdot)$ is a non-decreasing function of $t$. We have the following two lemmas regarding the functional $F(\cdot)$.

*Lemma 1* : $F(\cdot)$ is either a continuous monotonically decreasing function of $\nu_s$ or exhibits a discontinuity of the second kind at $\nu_{s,dc}$ where $t(\nu_{s,dc} < \eta_1) = \frac{1}{\bar{\alpha}}$. In addition, $\operatorname{sgn}(F(\cdot)) = \operatorname{sgn}(1-t\bar{\alpha})$.

*Proof* : The first property follows trivially from the fact that $F = \frac{1}{P_r}(-S + f(t))$, $-S, f$ both being decreasing functions of $\nu_s$ and $f(t)$ may be discontinuous. Solving for $-S(1-t\bar{\alpha})-(1-t) > 0$ results in $t > \left( \frac{1+S}{1+\bar{\alpha}S} \right)$ which is true for $\nu_s < \eta_1$ since $\left( e^{-R_1} \left( \frac{1+S}{1+\bar{\alpha}S} \right)^{\frac{1}{X-1}} \right) > 1, \nu_s < \eta_1$. Suppose $t(0,..) = e^{\frac{R_1}{1-X}} < \frac{1}{\bar{\alpha}}$, then using the fact that $t(\nu_s,..)$ is a decreasing function, $F(\cdot)$ will be a continuous, monotonically decreasing function. If, on the other hand, $t(0,..) = e^{\frac{R_1}{1-X}} > \frac{1}{\bar{\alpha}}$, then using the fact that $t(\eta_1,...) = e^{R_1} < \frac{1}{\bar{\alpha}}$, there exists $\nu_{s,dc} < \eta_1$ such that $t(\nu_{s,dc},..) = \frac{1}{\bar{\alpha}}$ at which $F(\cdot)$ exhibits a sign change $\lim_{\nu_s \to \nu_{s,dc}-} F(\nu_s,..) = -\infty, \lim_{\nu_s \to \nu_{s,dc}+} F(\nu_s,..) = \infty$ which is a second kind discontinuity $\square$.

*Lemma 2* : The probability intersection lies in the region $\max(\nu_{s,dc1}, \nu_{s,dc2}) \triangleq \nu_{s,dc} < \nu_s < \eta_1$ where

$$\nu_{s,dcX}=\begin{cases}-\frac{\chi-1}{P_s\left(\bar{\alpha}\chi-1\right)}, e^{\frac{R_1}{1-X}}>\frac{1}{\bar{\alpha}}, X=\left(X_1,X_2\right), e^{\frac{\left(-(X-1)\log(\bar{\alpha})+R_1\right)}{X}}\triangleq\chi \\ 0, else\end{cases} \quad (43)$$

and the evaluated probability bound is $\nu_r > F\left(..., X=\max\left(X_1,X_2\right)\right)$.

*Proof* : Assume without loss of generality that $X_1>X_2$. Note that $t(X)$ is an increasing function of $X$ for $\nu_s<\eta_1$, so $t_{X_1}>t_{X_2}$, and consequently $F(...,X_1)\triangleq F_{X_1}>F(...,X_2)\triangleq F_{X_2}$. Suppose both $F_{X_1},F_{X_2}$ are continuous, then $\nu_{s,dc1}=\nu_{s,dc2}=0$ and the statement holds since the intersection requires the higher of the two bounds. Assuming now both $F_{X_1},F_{X_2}$ are discontinuous, the intersection is void until both become positive, which happens for $\max\left(\nu_{s,dc1},\nu_{s,dc2}\right)<\nu_s<\eta_1$. In that region, $F_{X_1}>F_{X_2}$. Finally, assume one of the functions is continuous and the other is not. Surely, if $F_{X_2}$ is discontinuous then so will be $F_{X_1}$, and therefore only $F_{X_1}$ is discontinuous. The intersection is non-void for $\max\left(\nu_{s,dc1},\nu_{s,dc2}=0\right)<\nu_s<\eta_1$ and there $F_{X_1}>F_{X_2}$ as well □.

Next, the function $U\left(\nu_s,P_s,P_r,\alpha,\bar{\alpha},R_2,X\right)$ defined in (19) satisfies

$$\frac{\partial U}{\partial \nu_s}=-\frac{P_s}{P_r}+\left(\frac{X}{X-1}\right)\frac{P_s}{P_r}\left(Ze^{-R_2}\right)^{\frac{1}{X-1}}<0, \frac{\partial^2 U}{\partial^2 \nu_s}=e^{\frac{R_2}{1-X}}\frac{\bar{\alpha}P_s^2 X}{\left(X-1\right)^2}Z^{\frac{2-X}{X-1}}\geq 0 \quad (44)$$

and is increasing in $X$ for $\nu_s<\eta_2$. By this, the intersection (19) results in $\nu_r>U\left(...,X=\max\left(X_1,X_2\right)\right)$.Suppose $F\left(...,X\right)$ is continuous and therefore $\nu_{s,dc}=0$. If $F\left(0,...\right)\leq U\left(0,...\right)$ then using the facts that $F\left(\eta_1,...\right)=0<U\left(\eta_1,...\right)$, both functions being monotonically decreasing, and $U(\cdot)$ being convex $\cup$, there might be $n=2k, k=0,1..$ intersection points $\nu_{s,i}, i=1,2..n$ between the functions, such that

$$U\left(\nu_s,...\right)>F\left(\nu_s,...\right),\nu_s\in\bigcup_{k=0}^{\frac{n}{2}}\left[\nu_{s,2k},\nu_{s,2k+1}\right],U\left(\nu_s,...\right)<F\left(\nu_s,...\right),\nu_s\in\bigcup_{k=0}^{\frac{n}{2}-1}\left[\nu_{s,2k+1},\nu_{s,2(k+1)}\right],\nu_{s,0}=0,\nu_{s,n+1}=\eta_1. \quad (45)$$

It can be stated as a conjecture that $n=0,2$ are the only two values feasible for the number of intersection points. The conjecture holds for the MISO setting. For the opposite case of $F\left(0,...\right)\geq U\left(0,...\right)$, there might be $m$ intersection points $\nu_{s,i}, i=1,2..m$ between the functions where this time $m=2r+1, r=0,1...$ such that

$$U\left(\nu_s,...\right)<F\left(\nu_s,...\right),\nu_s\in\bigcup_{k=0}^{\frac{m-1}{2}}\left[\nu_{s,2k},\nu_{s,2k+1}\right],U\left(\nu_s,...\right)>F\left(\nu_s,...\right),\nu_s\in\bigcup_{k=0}^{\frac{m-1}{2}}\left[\nu_{s,2k+1},\nu_{s,2(k+1)}\right]. \quad (46)$$

For this case, the conjecture is that only $m=1$ is feasible, and this holds for the MISO as well. Finally, if $F\left(...,X\right)$ is discontinuous, recall that in the relevant interval of $\left[\max\left(\nu_{s,dc1},\nu_{s,dc2}\right)=\nu_{s,dc},\eta_1\right]$ the function is continuous, and $\lim_{\nu_s\to\nu_{s,dc}^+}F\left(\nu_s,...X\right)=\infty$, so the analysis is the same as for the continuous case, with the interval being $\left[\nu_{s,dc}\neq 0,\eta_1\right]$,as $F\left(\nu_{s,dc},...\right)>U\left(\nu_{s,dc},...\right)$. By using (45)-(46) we arrive to (22) where the three expressions stand for the three possible cases described and the most general form is given. Consider now the variable power allocation and the associated function with the derivatives

$$K(\nu_s,\alpha,\beta,P_s,P_r,X,R_1) \triangleq \frac{1}{P_r}\left(-S\left(\frac{1-t\bar{\alpha}}{1-t\bar{\beta}}\right)-\left(\frac{1-t}{1-t\bar{\beta}}\right)\right), \frac{\partial K}{\partial \nu_s}=\frac{\beta-\nu_s P_s(\alpha-\beta)}{(1-t\bar{\beta})^2}\frac{\partial t}{\partial \nu_s}-\frac{P_s(1-t\bar{\alpha})}{1-t\bar{\beta}}, \frac{\partial K}{\partial t}=\frac{\beta-\nu_s P_s(\alpha-\beta)}{(1-t\bar{\beta})^2}$$

From $\frac{\partial K}{\partial t}$ we see that for $\beta>\alpha$ the function increases with $t$ while for $\alpha>\beta$ the function is increasing for $\nu_s<\frac{\beta}{P_s(\alpha-\beta)}=\nu_s^*$ and decreasing for $\nu_s>\nu_s^*$. By lemma 1, $\operatorname{sgn}(K(\cdot))=\operatorname{sgn}(1-t\bar{\beta}),\nu_s\in[0,\eta_1]$. For a general $\beta$, the function $1-t\bar{\beta}$ may either be positive for all $\nu_s$, negative for all $\nu_s$ or change sign from negative to positive at some point. By limiting the relay's allocation coefficient to $\beta\ge\alpha$, the function behaves similarly to $1-t\bar{\alpha}$. This is since $1-e^{R_1}\bar{\beta}>1-e^{R_1}\bar{\alpha}>0$ so if there is a sign change at $\nu_{s,dcv}$, then $\nu_{s,dcv}\in[0,\eta_1]$,as $t$ is a decreasing function of $\nu_s$. Note that $K(t,\nu_s,...)$ is an increasing function of $t$. The following lemma establishes the dependence of $K(t,\nu_s,...)$ on $\nu_s$ for $\beta\ge\alpha$.

*Lemma 3* : $K(t,\nu_s,...)$ is a decreasing function of $\nu_s$ for $\left(\nu_s\middle|\operatorname{sgn}(1-t(\nu_s,..)\bar{\alpha})=\operatorname{sgn}(1-t(\nu_s,...)\bar{\beta})\right)$.

*Proof* : First, consider $-\left(\frac{1-t}{1-t\bar{\beta}}\right)$ which is decreasing . Next, $\frac{\partial}{\partial \nu_s}\left(\frac{1-t\bar{\alpha}}{1-t\bar{\beta}}\right)=\frac{(\alpha-\beta)}{(1-t\bar{\beta})^2}\frac{\partial t}{\partial \nu_s}>0$, so if $\left(\frac{1-t\bar{\alpha}}{1-t\bar{\beta}}\right)>0$ then $-S\left(\frac{1-t\bar{\alpha}}{1-t\bar{\beta}}\right)$ is decreasing □.

The fact that $K(t,\nu_s,...)$ is an increasing function of $t$ is sufficient to use the arguments derived for the $F(\cdot)$ function when evaluating the probability intersection with the substitution of $\nu_{s,dcv}$ instead of $\nu_{s,dc}$. The conjectures regarding the number of intersection points do not remain valid however. We also note that since $1-t\bar{\beta}>1-t\bar{\alpha}$, $K(\cdot)<F(\cdot)$, leading to higher decoding probabilities.

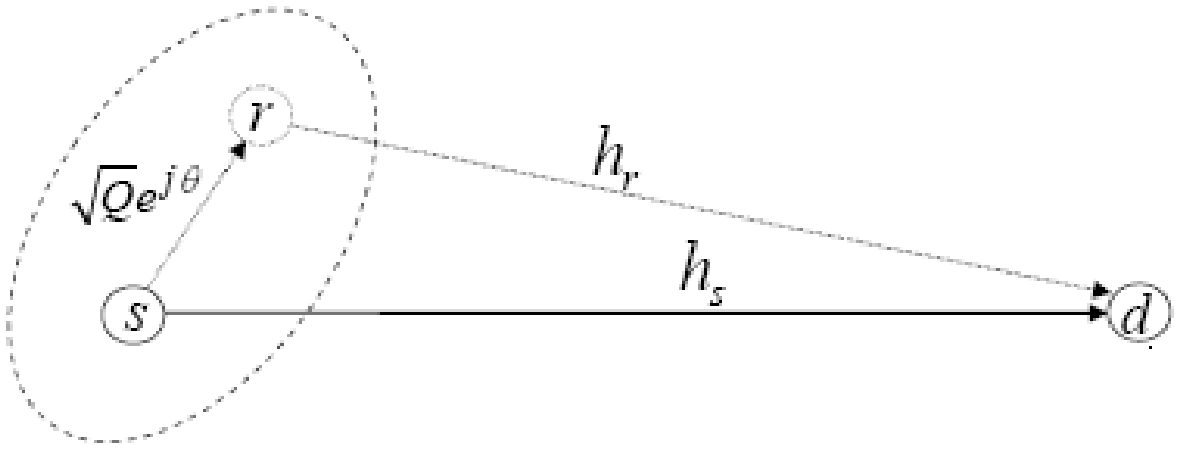

Fig. 1. A source-relay collocated network

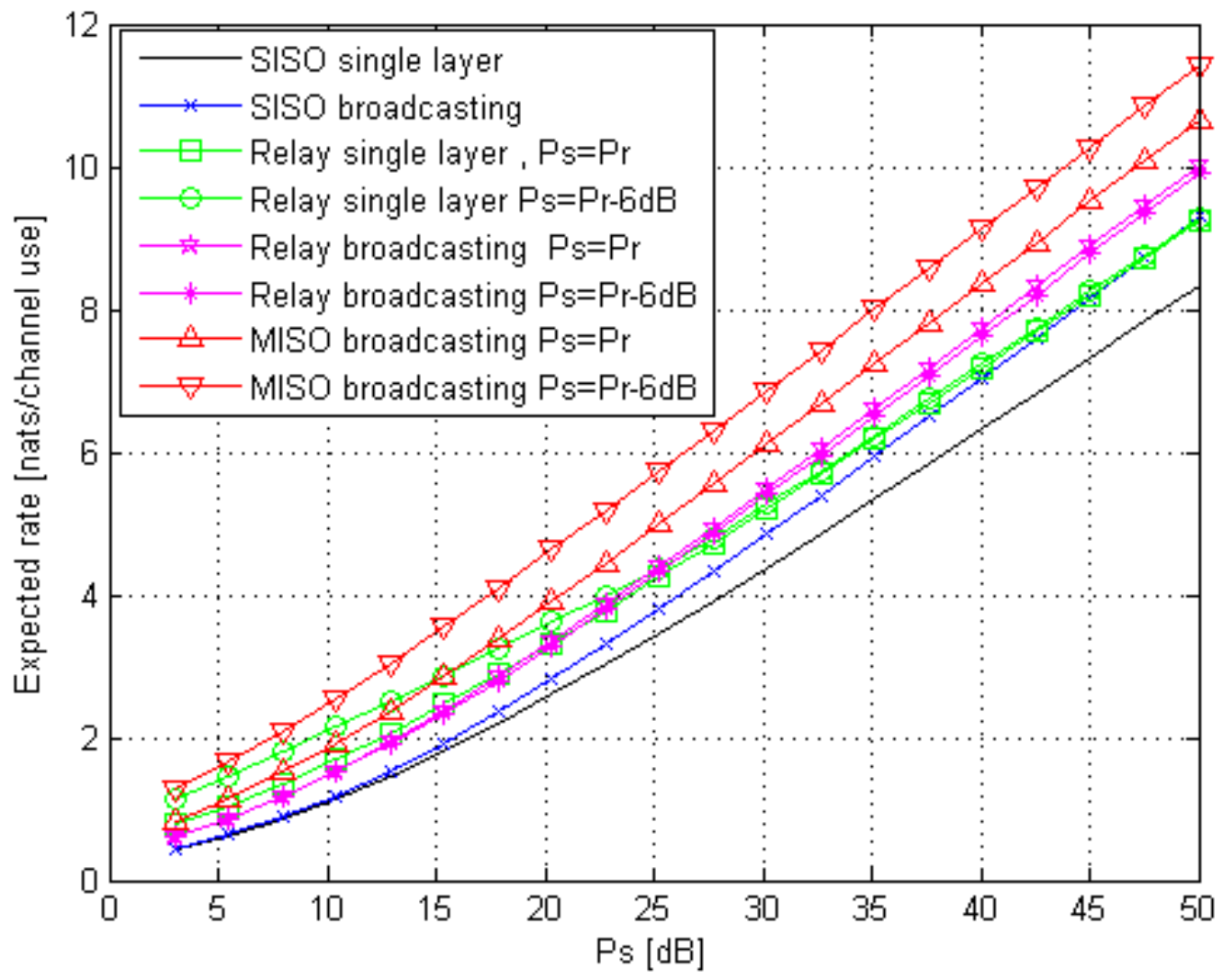

Fig. 2. Continuous broadcasting and single layer rates as a function of the source power and $\frac{P_s}{P_r}$ ratio. Continues broadcasting rates are lower bounds.

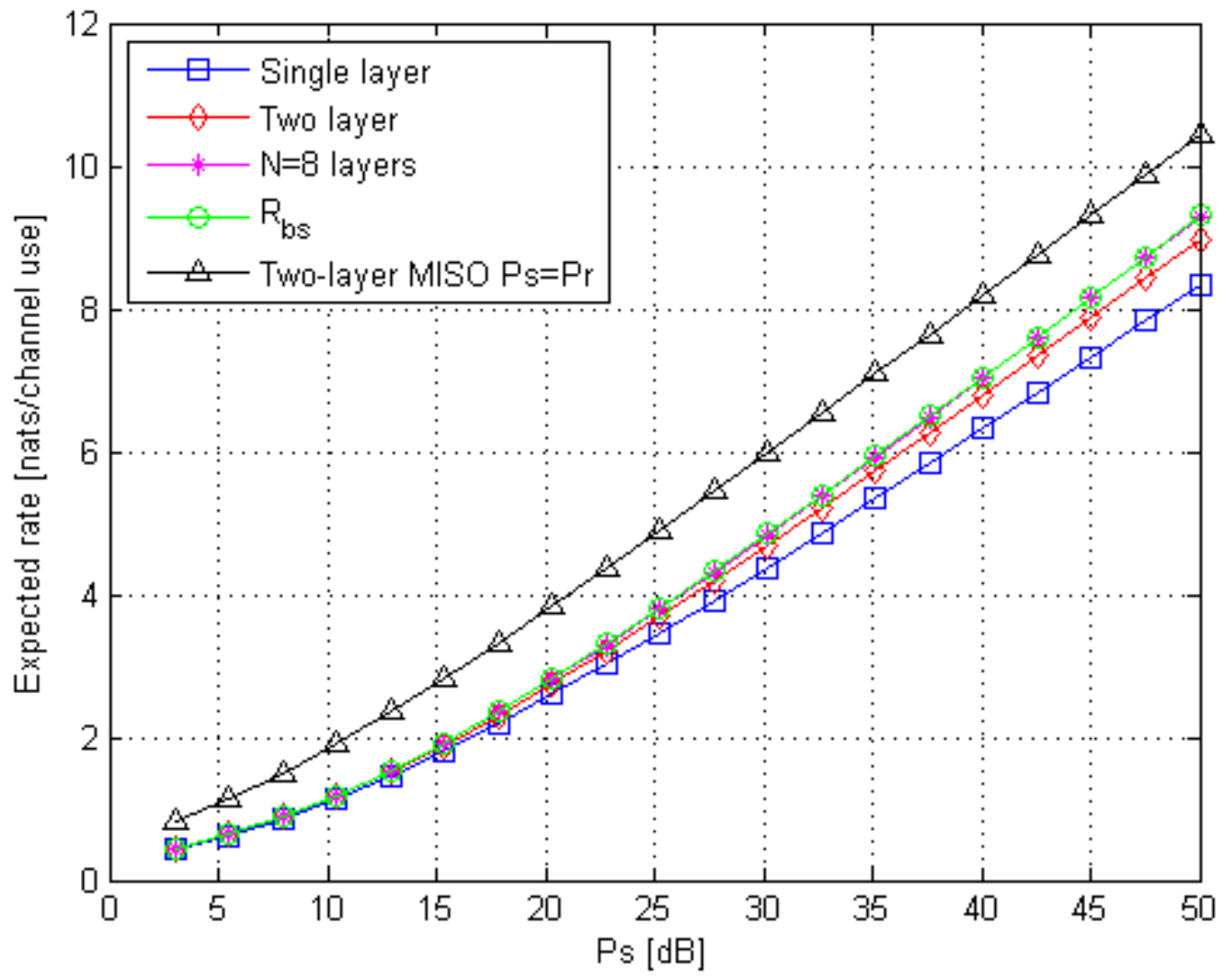

Fig. 3. Maximal achievable rates for the SISO (lower bound) single layer,two layer ,eight layer and continuous broadcasting computed from [4]. The range of improvement via the aid of a relay is bounded by the MISO rates of the appropriate transmission scheme ( $P_s = P_r$ for this example).

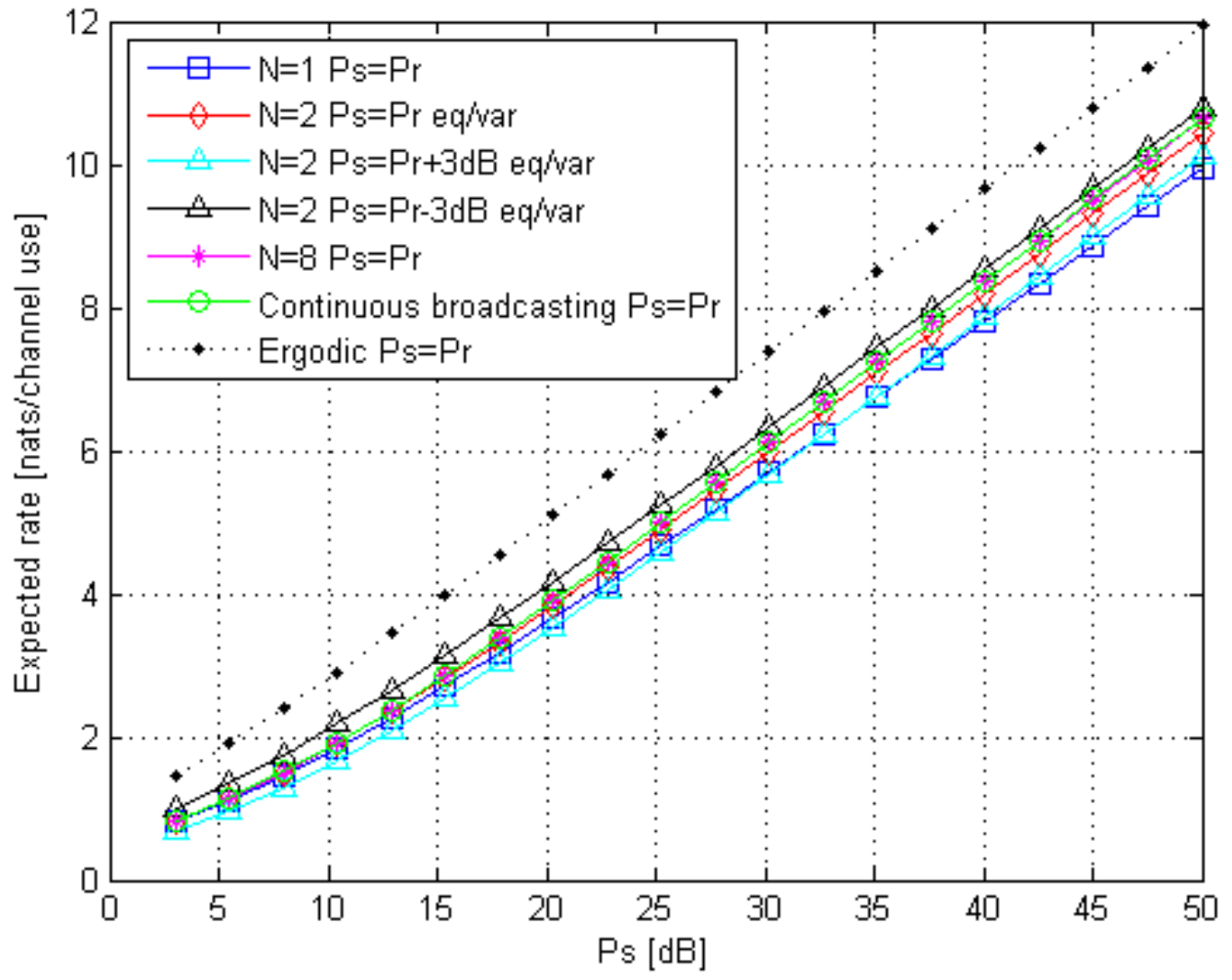

Fig. 4. Two layer 2x1 MISO achievable rates for the equal and unequal antenna layering power distributions as a function of the source power $P_s$ ,various $\frac{P_r}{P_s}$ ratios and number of layers $N = 1, 2, 8, \infty$ . The ergodic MISO presented as an upper bound.

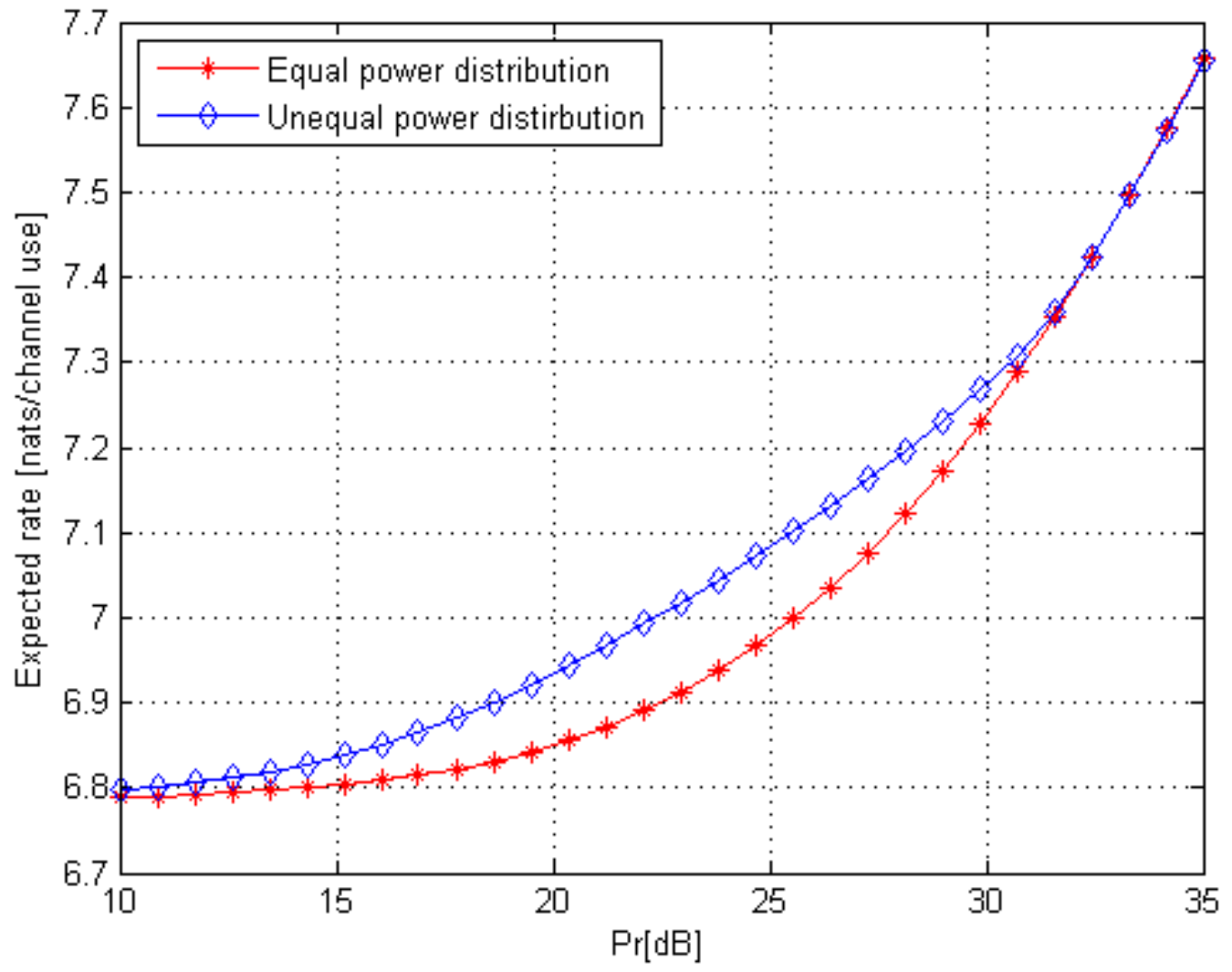

Fig. 5. Two layer 2x1 MISO achievable rates for an equal and unequal antenna layering power distributions as a function of the relay power $P_r$ , fixed source power of $P_s = 40\text{dB}$ .

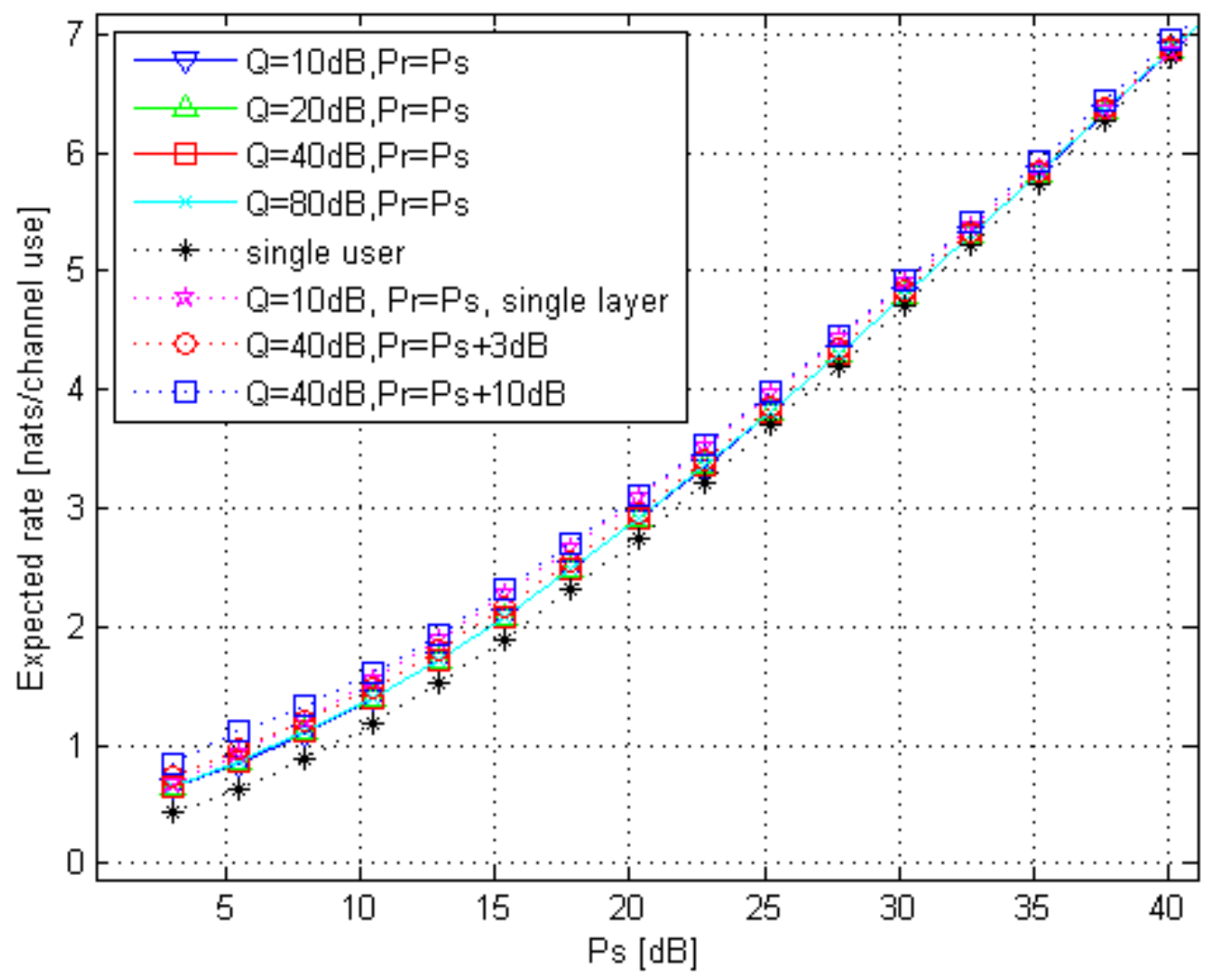

Fig. 6. Two layer oblivious SDF achievable rates for an equal antenna layering power distribution as a function of the source power $P_s$, collocation gain $Q$ and various $\frac{P_r}{P_s}$ ratios.

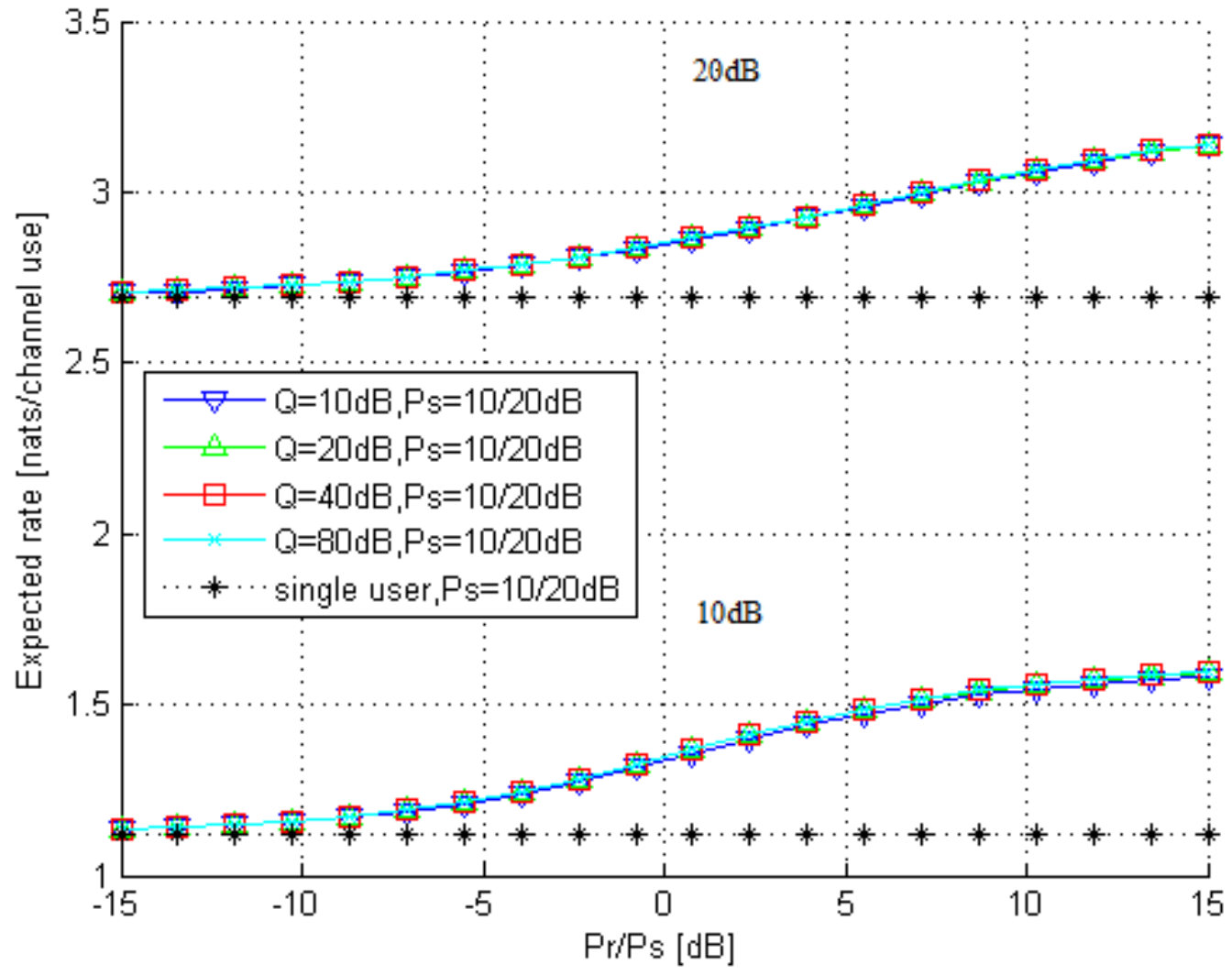

Fig. 7. Two layer oblivious SDF achievable rates for an equal antenna layering power distribution as a function of the $\frac{P_r}{P_s}$ ratio and the collocation gain $Q$, for a fixed source powers of $P_s = 10/20$dB. The SISO rate is a lower bound.

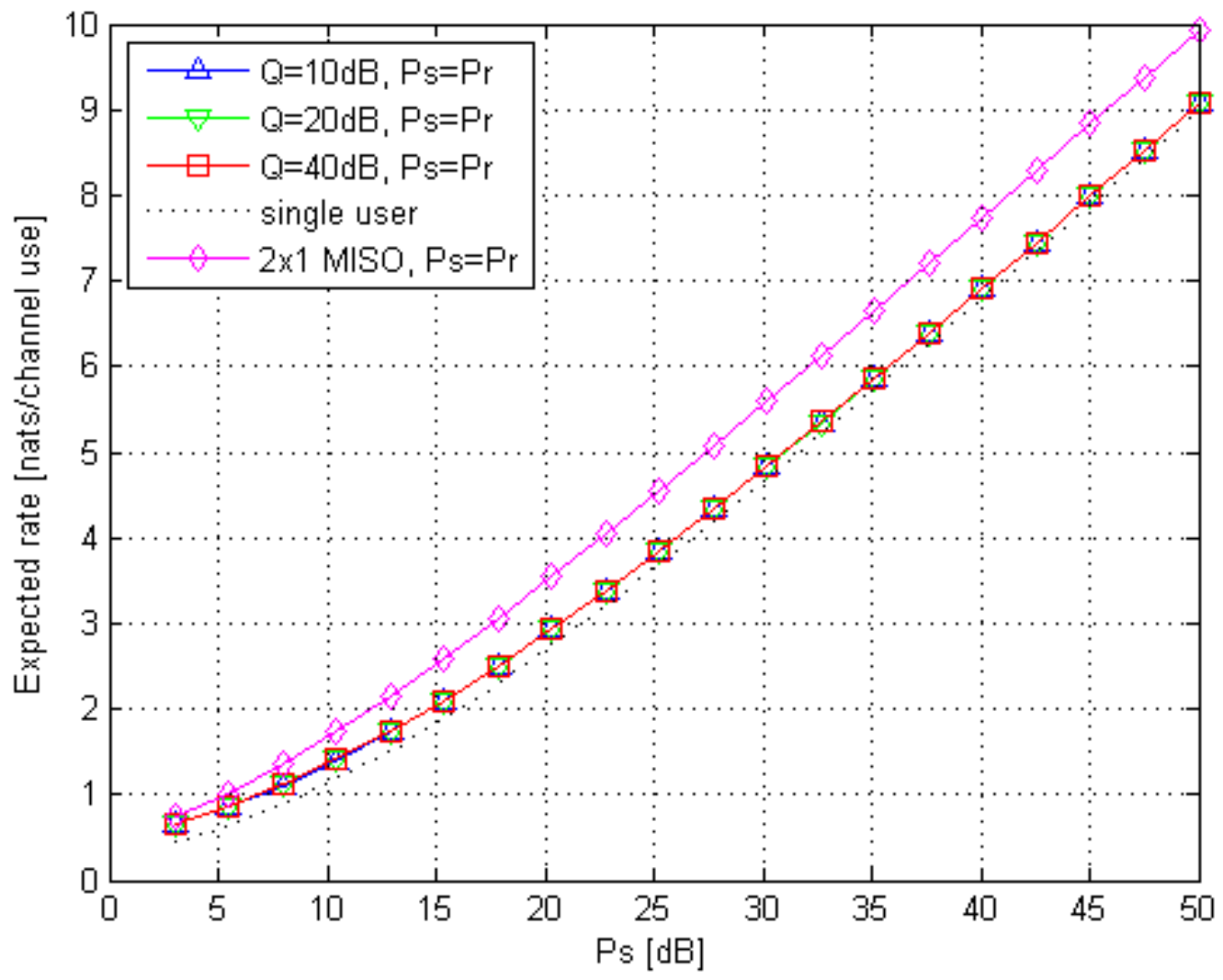

Fig. 8. Two layer oblivious SDF achievable rates for an unequal antenna layering power distribution as a function of the source power $P_s$, relay power $P_r = P_s$ and collocation gain $Q$ . The 2x1 MISO is an upper bound.

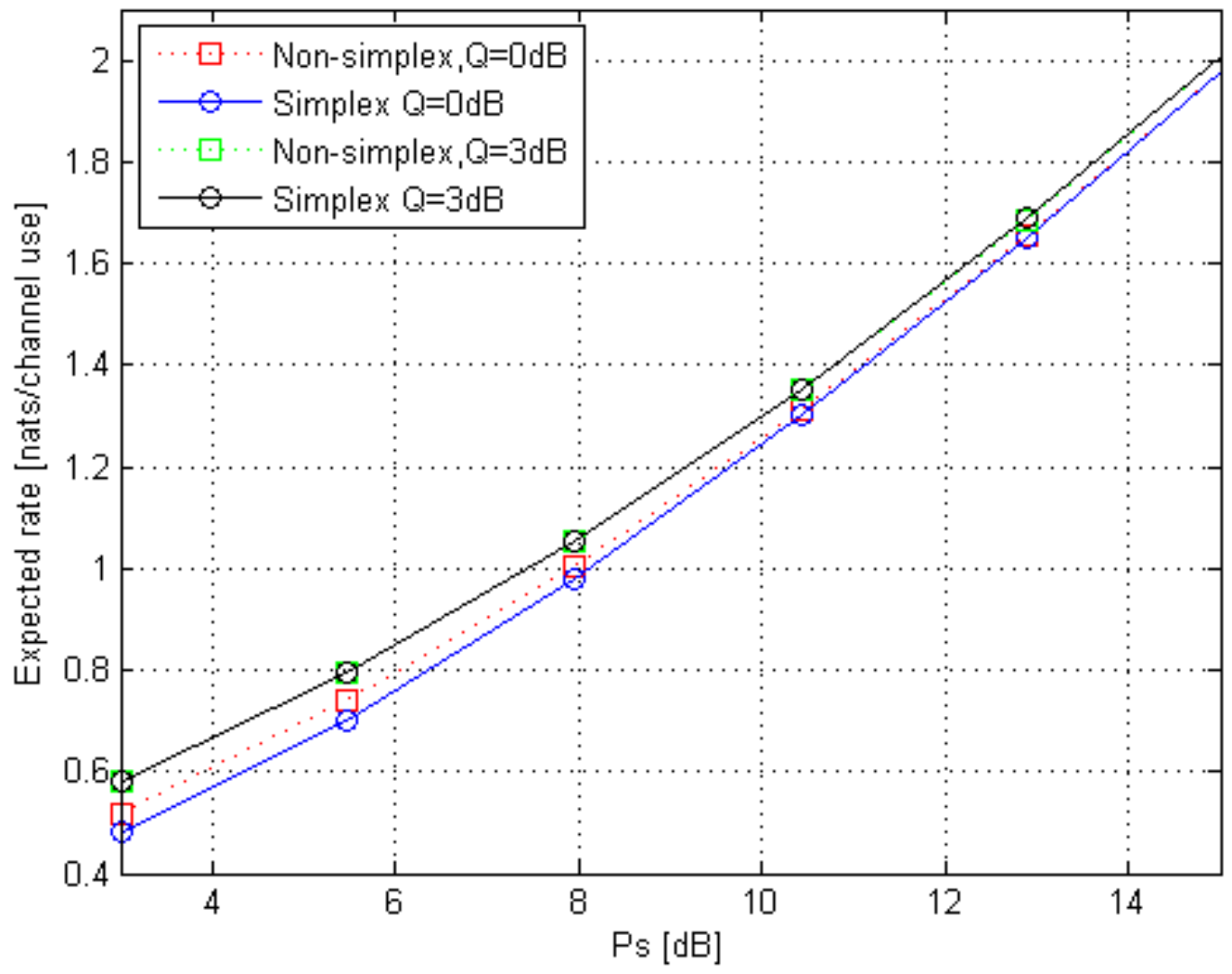

Fig. 9. Two layer oblivious SDF achievable rates for a full-duplex (non-simplex) and simplex relay under equal antenna layering power distribution , as a function of the source power $P_s$, relay power $P_r = P_s$ and collocation gain $Q$ .